\documentclass[11pt,a4paper]{article} 
\usepackage{jheppub} 

\usepackage{atlasphysics} 
\usepackage{subfigure}
\usepackage{mathrsfs}
\usepackage{bm}

\newcommand{\LUMI}{4.6~fb$^{-1}$ }
\def\mt{\ensuremath{m_{\mathrm{T}}}} 
\def\mtp2{\ensuremath{m_{\mathrm{T}2}^{H}}}

\usepackage{preprintcover}  

\PreprintCoverPaperTitle{\boldmath Search for charged Higgs bosons through the violation of lepton universality in $t\bar{t}$ events using $pp$ collision data at $\sqrt{s} = 7~\mbox{TeV}$ with the ATLAS experiment} 

\PreprintIdNumber{CERN-PH-EP-2012-347}  

\PreprintCoverAbstract{
In several extensions of the Standard Model, the top quark can 
decay into a bottom quark and a light charged Higgs boson $H^+$, 
$t \rightarrow bH^+$, in addition to the Standard Model decay 
$t \rightarrow bW$. Since $W$ bosons decay to the three lepton 
generations equally, while $H^+$ may predominantly decay into 
$\tau\nu$, charged Higgs bosons can be searched for using the 
violation of lepton universality in top quark decays. The 
analysis in this paper is based on \LUMI of proton--proton 
collision data at $\sqrt{s} = 7~\mbox{TeV}$ collected by 
the ATLAS experiment at the Large Hadron Collider. 
Signatures containing leptons ($e$ or $\mu$) and/or a 
hadronically decaying $\tau$ ($\tau_{\text{had}}$) are 
used. Event yield ratios between $e+\tau_{\text{had}}$ 
and $e+\mu$, as well as between $\mu+\tau_{\text{had}}$ 
and $\mu+e$, final states are measured in the data and 
compared to predictions from simulations. This ratio-based 
method reduces the impact of systematic uncertainties in 
the analysis. No significant deviation from the Standard 
Model predictions is observed. With the assumption that 
the branching fraction ${\cal B}(H^+ \rightarrow \tau\nu)$ 
is 100\%, upper limits in the range 3.2\%--4.4\% can be placed 
on the branching fraction ${\cal B}(t\rightarrow bH^+)$ for 
charged Higgs boson masses $m_{H^+}$ in the range 90--140~GeV. 
After combination with results from a search for charged Higgs 
bosons in $t\bar{t}$ decays using the $\tau_{\text{had}}$+jets 
final state, upper limits on ${\cal B}(t\rightarrow bH^+)$ can 
be set in the range 0.8\%--3.4\%, for $m_{H^+}$ in the range 
90--160~GeV. 
}  

\PreprintJournalName{JHEP}  

\title{\boldmath Search for charged Higgs bosons through the violation of lepton universality in $t\bar{t}$ events using $pp$ collision data at $\sqrt{s} = 7~\mbox{TeV}$ with the ATLAS experiment}

\author{The ATLAS Collaboration}

\date{\today}

\abstract{ 
In several extensions of the Standard Model, the top quark can 
decay into a bottom quark and a light charged Higgs boson $H^+$, 
$t \rightarrow bH^+$, in addition to the Standard Model decay 
$t \rightarrow bW$. Since $W$ bosons decay to the three lepton 
generations equally, while $H^+$ may predominantly decay into 
$\tau\nu$, charged Higgs bosons can be searched for using the 
violation of lepton universality in top quark decays. The 
analysis in this paper is based on \LUMI of proton--proton 
collision data at $\sqrt{s} = 7~\mbox{TeV}$ collected by 
the ATLAS experiment at the Large Hadron Collider. 
Signatures containing leptons ($e$ or $\mu$) and/or a 
hadronically decaying $\tau$ ($\tau_{\text{had}}$) are 
used. Event yield ratios between $e+\tau_{\text{had}}$ 
and $e+\mu$, as well as between $\mu+\tau_{\text{had}}$ 
and $\mu+e$, final states are measured in the data and 
compared to predictions from simulations. This ratio-based 
method reduces the impact of systematic uncertainties in 
the analysis. No significant deviation from the Standard 
Model predictions is observed. With the assumption that 
the branching fraction ${\cal B}(H^+ \rightarrow \tau\nu)$ 
is 100\%, upper limits in the range 3.2\%--4.4\% can be placed 
on the branching fraction ${\cal B}(t\rightarrow bH^+)$ for 
charged Higgs boson masses $m_{H^+}$ in the range 90--140~GeV. 
After combination with results from a search for charged Higgs 
bosons in $t\bar{t}$ decays using the $\tau_{\text{had}}$+jets 
final state, upper limits on ${\cal B}(t\rightarrow bH^+)$ can 
be set in the range 0.8\%--3.4\%, for $m_{H^+}$ in the range 
90--160~GeV. 

}

\begin{document}

\maketitle


\section{Introduction}

Several non-minimal Higgs scenarios, e.g.\ Two Higgs Doublet 
Models (2HDM)~\cite{Lee:1973iz} predict the existence of charged 
Higgs bosons ($H^+$ and $H^-$).\footnote{In the following, charged 
Higgs bosons are denoted $H^+$, with the charge-conjugate $H^-$ 
always implied. Hence, $\tau$ denotes a positively charged $\tau$ 
lepton.} Their observation would clearly indicate physics 
beyond the Standard Model (SM), because this theory contains no 
elementary charged scalar particle. In several models, e.g.\ 
a type-II 2HDM describing the Higgs sector of the Minimal 
Supersymmetric extension of the Standard Model (MSSM)~\cite{Fayet:1976et,Fayet:1977yc,Farrar:1978xj,Fayet:1979sa,Dimopoulos:1981zb}, 
the main production mode for charged Higgs bosons with a mass $m_{H^+}$ 
smaller than the top quark mass ($m_{\text{top}}$) is through top quark 
decays $t \rightarrow bH^+$. The dominant source of top quarks at the 
Large Hadron Collider (LHC) is through $t\bar{t}$ 
production.\footnote{Since the cross section for $H^+$ production 
from events containing a single top quark is much smaller, this 
production mode is not considered here.}

The combined LEP lower limit on the charged Higgs boson mass is about 
90~GeV~\cite{:2001xy}. Results from direct searches for charged Higgs 
bosons decaying via $H^+ \rightarrow \tau\nu$ using 4.6~fb$^{-1}$ 
of LHC data were recently presented by the ATLAS 
collaboration~\cite{ourpaper}, with upper limits on the branching 
fraction ${\cal B}(t \rightarrow bH^{+})$ between 5\% and 1\% for 
charged Higgs boson masses ranging from 90~GeV to 160~GeV, respectively. 
Using about 2~fb$^{-1}$ of LHC data, the CMS collaboration established 
upper limits on ${\cal B}(t \rightarrow bH^{+})$ in the range 4--2\% for 
charged Higgs boson masses between 80~GeV to 160~GeV~\cite{cmshpluspaper}. 
In all of these measurements, as well as in this paper (unless otherwise 
specified), the assumption ${\cal B}(H^+ \rightarrow \tau\nu) = 100\%$ 
is made.

This paper uses an alternative technique~\cite{addzero} for $H^+$ searches 
in the mass range 90--160~GeV. Instead of using the shape of discriminating 
variables in order to search for a local excess of events above the predicted 
SM background, this analysis is based on the measurement of a ratio of event 
yields between two $t\bar{t}$ final states, which in turn allows for the 
cancellation of most of the systematic uncertainties. In top quark decays, 
$W$ bosons decay equally to leptons of the three generations, while $H^+$ 
may decay predominantly into $\tau\nu$. Hence, an excess of $t\bar{t}$ 
events with at least one hadronically decaying $\tau$ lepton 
($\tau_{\text{had}}$) in the final state, as compared to the 
rate for $t\bar{t}$ events with only electrons and/or muons, is 
a signature for charged Higgs bosons. A measurement of event yield 
ratios $R_l$ for $t\bar{t} \to b\bar{b} + l\tau_{\text{had}} + N\nu$ 
and $t\bar{t} \to b\bar{b} + ll' + N\nu$, where $ N\nu$ stands for any 
number of neutrinos and where $l$ and $l'$ are electrons and muons, with 
$l \neq l'$, is performed:
\begin{equation}
R_l = {
{\cal B}(t\bar{t} \to b\bar{b} + l\tau_{\text{had}} + N\nu) 
\over 
{\cal B}(t\bar{t} \to b\bar{b} + ll' + N\nu)
}. 
\end{equation}

This study is performed in a model-independent way, and so exclusion limits 
are given in terms of ${\cal B}(t\rightarrow bH^+)$, as well as in the 
$m_h^{\mathrm{max}}$ scenario~\cite{Carena:2002qg} of the MSSM. 
The results are based on \LUMI of data from $pp$ collisions at 
$\sqrt{s}=7~\mbox{TeV}$, collected in 2011 with the ATLAS 
experiment~\cite{Aad:2008zzm} at the LHC. These data, as well as 
the simulated samples used in the analysis, are the same as in 
ref.~\cite{ourpaper} and are described briefly in section~2. Then, 
in section~3, an event selection aimed at collecting a data sample 
enriched in $t\bar{t}$ events is presented, together with the data-driven 
methods to estimate the backgrounds due to misidentified electrons, muons 
and hadronically decaying $\tau$ leptons. Exclusion limits in terms of 
${\cal B}(t\rightarrow bH^+)$ and $\tan\beta$ are discussed in section~4, 
based on the measured ratios among event yields in $\tau_{\text{had}}$+lepton 
and dilepton final states. Finally, a summary is given in section~5.


\section{ATLAS data and simulated events}

The ATLAS detector \cite{Aad:2008zzm} consists of an inner tracking 
detector with coverage in pseudorapidity\footnote{
ATLAS uses a right-handed coordinate system with
its origin at the nominal interaction point (IP) in the centre
of the detector and the $z$-axis along the beam pipe. The $x$-axis
points from the IP to the centre of the LHC ring, and the $y$-axis
points upwards. Cylindrical coordinates $(r,\phi)$ are used in the
transverse plane, $\phi$ being the azimuthal angle around the beam pipe.
The pseudorapidity is defined in terms of the polar angle $\theta$ as
$\eta=-\ln\tan(\theta/2)$.}
up to $|\eta|=2.5$, surrounded by a thin 2~T superconducting solenoid, 
a calorimeter system extending up to $|\eta|=4.9$ for the detection of 
electrons, photons and hadronic jets, and a large muon spectrometer 
extending up to $|\eta|=2.7$ that measures the deflection of muon 
tracks in the field of three superconducting toroid magnets. A 
three-level trigger system is used, which reduces the recorded 
event rate to about 300~Hz.

In ATLAS, electrons are reconstructed by matching clustered 
energy deposits in the electromagnetic calorimeter to tracks 
reconstructed in the inner detector, and muons are required 
to contain matching inner detector and muon spectrometer tracks. 
The combination of all sub-systems provides precise lepton 
measurements in the pseudorapidity range $|\eta|< 2.5$. 
Jets, as well as the magnitude of the missing transverse momentum, 
$\met$, are reconstructed using energy deposits over the full 
coverage of the calorimeters, up to $|\eta|=4.9$. The anti-$k_t$ 
algorithm~\cite{Cacciari:2008gp, Cacciari:2005hq} with a radius 
parameter value of \mbox{$R = 0.4$} is used for jet reconstruction. 
At least 75\% of the tracks associated to a jet (weighted by 
their transverse momenta) must point to the primary vertex, 
corresponding to the hardest interaction. This 
requirement on the ``Jet Vertex Fraction''~\cite{Abazov:2006yb} 
allows the identification of jets originating from the hard-scatter 
interaction. An algorithm combining impact-parameter information 
with the explicit observation of a secondary vertex~\cite{sys_b} 
is used in order to identify jets initiated by $b$-quarks, within 
$|\eta| < 2.4$. The working point chosen for this study corresponds 
to an average efficiency of about $70\%$ for $b$-jets with 
$\pt >20 \GeV$ in $t\bar{t}$ events and a rejection factor of 
about 130 for light-quark jets. In order to reconstruct hadronically 
decaying $\tau$ leptons, anti-$k_t$ jets with either one or three 
associated tracks, depositing $\ET > 10 \GeV$ in the calorimeter, 
are considered as $\tau$ candidates~\cite{taus}. Dedicated algorithms 
are used to reject electrons and muons. The $\tau$ candidates are further 
required to have a visible transverse momentum $\pt^{\tau} >20 \GeV$ 
and to be within $|\eta| < 2.3$. The hadronic $\tau$ decays are 
identified using a likelihood criterion designed to discriminate against 
quark- and gluon-initiated jets. The working point chosen for this study 
corresponds to an efficiency of about 30\% for hadronically decaying 
$\tau$ leptons with $\pt^{\tau} > 20 \GeV$ in $Z \rightarrow \tau\tau$ events, 
leading to a rejection factor of about 100--1000 for jets. Selected 
$\tau$ candidates fulfilling the identification criteria are referred 
to as ``$\tau$ jets''.
When objects selected using the criteria above overlap geometrically, 
the following procedures are applied, in this order: muons are rejected 
if found within $\Delta R = \sqrt{\Delta\eta^2 + \Delta\phi^2} < 0.4$ of 
any jet with $\pt > 25 \GeV$; a $\tau$ candidate is rejected if found 
within $\Delta R < 0.4$ of a $b$-tagged jet, or within $\Delta R < 0.2$ 
of a selected muon or electron; jets are removed if they are within 
$\Delta R < 0.2$ of a selected $\tau$ jet or electron.

The same full 2011 data set and simulated samples as the analysis 
in ref.~\cite{ourpaper} are used, corresponding to an integrated 
luminosity of 4.6~fb$^{-1}$, with an uncertainty of 
3.9\%~\cite{lumi,Aad:2011dr}. 
In addition to the SM pair production and decay of top quarks, 
$t\bar{t} \rightarrow b\bar{b}W^+W^-$,\footnote{In the simulated SM 
$t\bar{t}$ events, all leptonic $W$ decay modes have the same branching 
fraction (10.8\%).} the background processes include the production of 
single top quark, $W$+jets, $Z/\gamma^*$+jets, diboson, and multi-jet 
events. Except for the last, which is estimated using data-driven methods, 
the SM backgrounds are determined using the simulated samples summarised in 
table~\ref{xs}. 
In addition to the SM background samples, three types 
of signal samples are produced with PYTHIA~6.425~\cite{Sjostrand:2006za} 
for $90\GeV < m_{H^+} < 160\GeV$: 
$t\bar{t} \rightarrow b\bar{b}H^+W^-$, 
$t\bar{t} \rightarrow b\bar{b}H^-W^+$ and 
$t\bar{t} \rightarrow b\bar{b}H^+H^-$, where  
charged Higgs bosons decay as $H^+ \rightarrow \tau\nu$. 
TAUOLA~1.20~\cite{Was:2004dg} is used for $\tau$ decays, and 
PHOTOS~2.15~\cite{Barberio:1990ms} is used for photon radiation 
from charged leptons. All generated events are propagated through 
a detailed GEANT4 simulation~\cite{Agostinelli:2002hh,simu}, 
and they are reconstructed using the same algorithms as the data.

\begin{table}[h!]
  \begin{center}
    \begin{tabular}{l|ll}
Process   &  Generator & Cross section [pb] \\ 
\hline \hline
SM $t\bar{t}$ with at least one lepton $\ell = e,\,\mu,\,\tau$ & 
MC@NLO~4.01~\cite{Frixione:2002ik} & 
91~~~~~~~~~~\cite{hathor} \\ 
\hline
Single top quark $t$-channel (with $\ell$) & 
AcerMC~3.8~~~~\;\cite{acer} &
21~~~~~~~~~~\cite{Kidonakis:2011wy} \\ 
Single top quark $s$-channel (with $\ell$) & 
MC@NLO~4.01~\cite{Frixione:2002ik} &
$1.5\phantom{\times 10^0}$~\,~\cite{Kidonakis:2010tc} \\
Single top quark $Wt$-channel (inclusive) & 
MC@NLO~4.01~\cite{Frixione:2002ik} & 
16~~~~~~~~~~\cite{Kidonakis:2010ux} \\ 
\hline
$W \rightarrow \ell\nu$ & 
ALPGEN~~2.13~\cite{Mangano:2002ea} & 
$3.1 \times 10^4$~\cite{Gavin:2012kw} \\ 
\hline
$Z/\gamma^* \rightarrow \ell\ell$ with $m(\ell\ell) > 10~\mbox{GeV}$ & 
ALPGEN~~2.13~\cite{Mangano:2002ea} & 
$1.5 \times 10^4$~\cite{Gavin:2010az} \\
\hline
$WW$  & HERWIG~6.520~\cite{hw} & 
17~~~~~~~~~~\cite{Campbell:2011bn} \\ 
$ZZ$  & HERWIG~6.520~\cite{hw} & 
$1.3\phantom{\times 10^0}$~\,~\cite{Campbell:2011bn} \\
$WZ$  & HERWIG~6.520~\cite{hw} & 
$5.5\phantom{\times 10^0}$~\,~\cite{Campbell:2011bn} \\
\hline
$H^+$ signal with ${\cal B}(t \to bH^+)=3\%$ &
PYTHIA~6.425~\,\cite{Sjostrand:2006za} & 9.9 \\
    \end{tabular}
\caption{Cross sections for the simulated processes and the generators used to model them. All background cross sections are normalised to next-to-next-to-leading-order (NNLO) predictions, except for diboson event production where the next-to-leading-order prediction (NLO) is used. For the diboson events, a filter is applied at the generator level, by requiring at least one electron or muon with $\pt > 10~\mbox{GeV}$ and $|\eta|<2.8$.
\label{xs}}
  \end{center}
\end{table}


\section{Event selection and background determination}
\label{EventSel}

This analysis uses events passing a single-lepton trigger with an $\et$ 
threshold of 20~GeV or 22~GeV for electrons and a $\pt$ threshold of 18~GeV 
for muons. In order to select a sample enriched in $t\bar{t}$ events, the 
following requirements are made:
\begin{itemize}
\item one charged lepton $l$ ($e,\,\mu$) having 
$E_{\text{T}} > 25~\mbox{GeV}$ ($e$) or 
$\pt > 25~\mbox{GeV}$ ($\mu$) and matched to the
corresponding trigger object;
\item at least two jets having $\pt > 20~\mbox{GeV}$ and $|\eta|<2.4$, 
including exactly two $b$-tags;
\item either exactly one $\tau$ jet with $\pt^{\tau} > 25~\mbox{GeV}$ and 
$|\eta|<2.3$ with no additional charged lepton, or exactly one additional 
charged lepton $l'$ with $E_{\text{T}}$ or $\pt$ above 25~GeV and a 
different flavour than the trigger-matched lepton;
\item $\met > 40~\mbox{GeV}$.
\end{itemize}

At this stage, the selected events are classified into two categories 
according to the single-lepton trigger that they fire: 
an electron trigger (EL) or a muon trigger (MU). Each category 
contains $\tau_{\text{had}}$+lepton and dilepton ($ll'$) events. 
The lepton appearing first in the final state 
is, by convention, matched to the corresponding trigger object. 
The EL category therefore consists of $e+\tau_{\text{had}}$ and 
$e+\mu$ events, while the MU category contains $\mu+\tau_{\text{had}}$ 
and $\mu+e$ events. Events firing both a single-electron trigger and a 
single-muon trigger are assigned to both categories, and accounted for 
in the combined limit setting, see section~\ref{sec:limit}.

The analysis uses the generalised transverse mass 
$m_{\mathrm{T2}}^H$~\cite{Gross:2009wia} as a selection variable. 
By construction, it gives an event-by-event lower bound on the mass 
of the charged ($W$ or Higgs) boson produced in the top quark decay. 
Hence, it is larger than the true charged Higgs boson mass $m_{H^+}$ 
and smaller than $m_{\text{top}}$. For incorrect pairings of $\tau$ 
jets or leptons with $b$-jets, the numerical determination of 
$m_{\mathrm{T2}}^H$ may fail, hence only events with 
$m_{\mathrm{T2}}^H>0$ are kept in the following.

\subsection{Backgrounds due to misidentified electrons and muons}
\label{sec:lepfakes}

A significant background for the search described in this paper consists of 
events with reconstructed electrons and muons arising from the semileptonic 
decay of hadrons with $b$- or $c$-quarks, from the decay-in-flight of 
$\pi$ or $K$ mesons and, in the case of electrons, from $\pi^{0}$ mesons, 
photon conversions or shower fluctuations. These are referred to as 
``misidentified leptons'' in the following. Two data samples are defined, 
which differ only in the lepton identification criteria. The \emph{tight} 
sample corresponds to the selection used in the analysis and contains 
mostly events with real leptons. The \emph{loose} sample is obtained 
by loosening the isolation and identification requirements, and it 
contains mostly events with misidentified leptons.\footnote{By construction, 
the tight sample is  a subset of the loose sample.} The efficiencies 
$p_{\text{r}}$ and $p_{\text{m}}$ for a real or misidentified lepton, 
respectively, to be detected as a tight lepton, are determined from 
data, with the same method as in ref.~\cite{ourpaper}. In the final 
parameterisation of $p_{\text{r}}$ and $p_{\text{m}}$, dependencies 
on the pseudorapidity of the lepton, its distance $\Delta R$ to the 
nearest jet and the leading jet $\pt$ are taken into account. Based 
on these efficiencies, the number of misidentified leptons passing 
the final requirements can be calculated by weighting each event in 
the data sample with one loose lepton, according to the following 
per-lepton weights $w_{l}$:
\begin{itemize}
\item for a loose but not tight lepton, 
$w_{l\text{L}} = \displaystyle{{p_{\text{m}}p_{\text{r}} \over (p_{\text{r}} - p_{\text{m}})}}$;
\item for a tight lepton, 
$w_{l\text{T}} = \displaystyle{{p_{\text{m}}(p_{\text{r}}-1) \over (p_{\text{r}} - p_{\text{m}})}}$.
\end{itemize}

\subsection{Backgrounds due to misidentified $\tau$ jets}
\label{JettoTauFakes}

About 51\% of the simulated $t\bar{t}$ events in the 
$\tau_{\text{had}}$+lepton final state contain a $\tau$ 
jet matched to a hadronically decaying $\tau$ lepton at 
the generator level. In the other events, the $\tau$ 
jet is called ``misidentified''. It originates from 
leptons ($e$, $\mu$) in 3\% of the simulated events 
and hadronic objects (initiated by light quarks, 
$b$-quarks or gluons) in 46\%. Data-driven methods are 
used in order to determine the probability of misidentification 
from electrons and hadronic jets. In the case of electrons, the 
misidentification probabilities are measured using a $Z \to ee$ 
control region in the data~\cite{taus} and then applied to the 
simulated events, as in the analysis in ref.~\cite{ourpaper}.
The majority of misidentified $\tau$ jets in the final event selection 
originate from jets, for which the misidentification probability depends 
on the initial parton (light quark, heavy-flavour quark or gluon). All 
jet types occur in $t\bar{t}$ events, and it is not possible to accurately 
predict the fraction of each of them, potentially leading to a large 
systematic uncertainty on the jet$\,\to\tau_{\text{had}}$ misidentification 
probability. However, the influence of all jet types other than light-quark 
jets can effectively be eliminated by categorising all events in terms 
of the charge of the lepton relative to the $\tau$ jet as 
opposite-sign (OS) or same-sign (SS) events. All processes with 
gluon and $b$-quark jets produce positively and negatively charged 
misidentified $\tau$ objects at the same rate. On the other hand, 
the light-quark jet component in SS events represents both charge 
misreconstruction and quarks which fragment such that the leading 
charged particle does not have the same charge as the initial quark. 
Giving a negative weight to the SS events therefore cancels, on average, 
the gluon and heavy-flavour-quark jet contributions from the OS events, 
leaving only light-quark jets misidentified as $\tau$ jets. 

\begin{figure}[h!]
\begin{center}
\subfigure[]{
\includegraphics[width=0.38\textwidth]{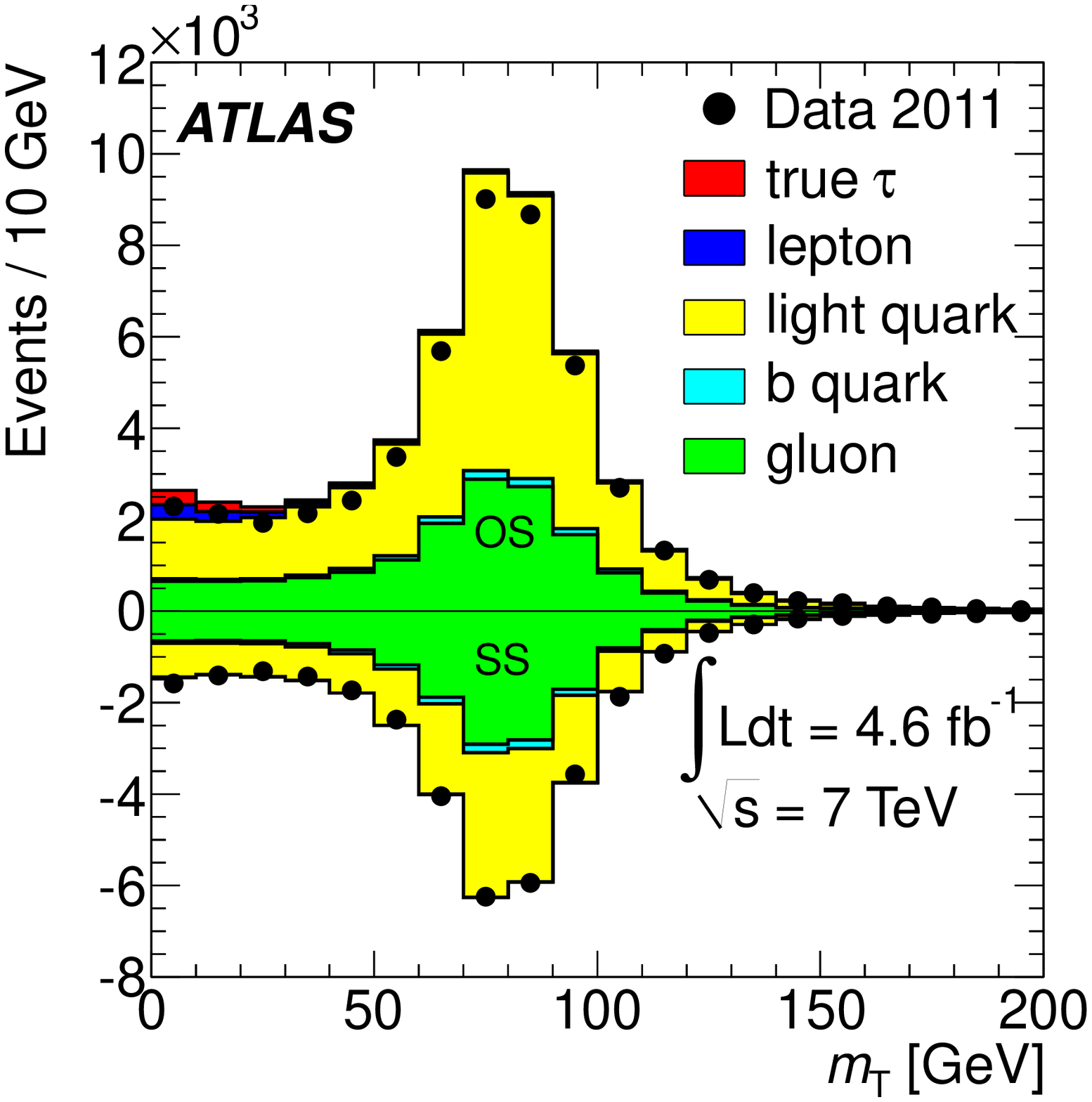}
}
\subfigure[]{
\includegraphics[width=0.38\textwidth]{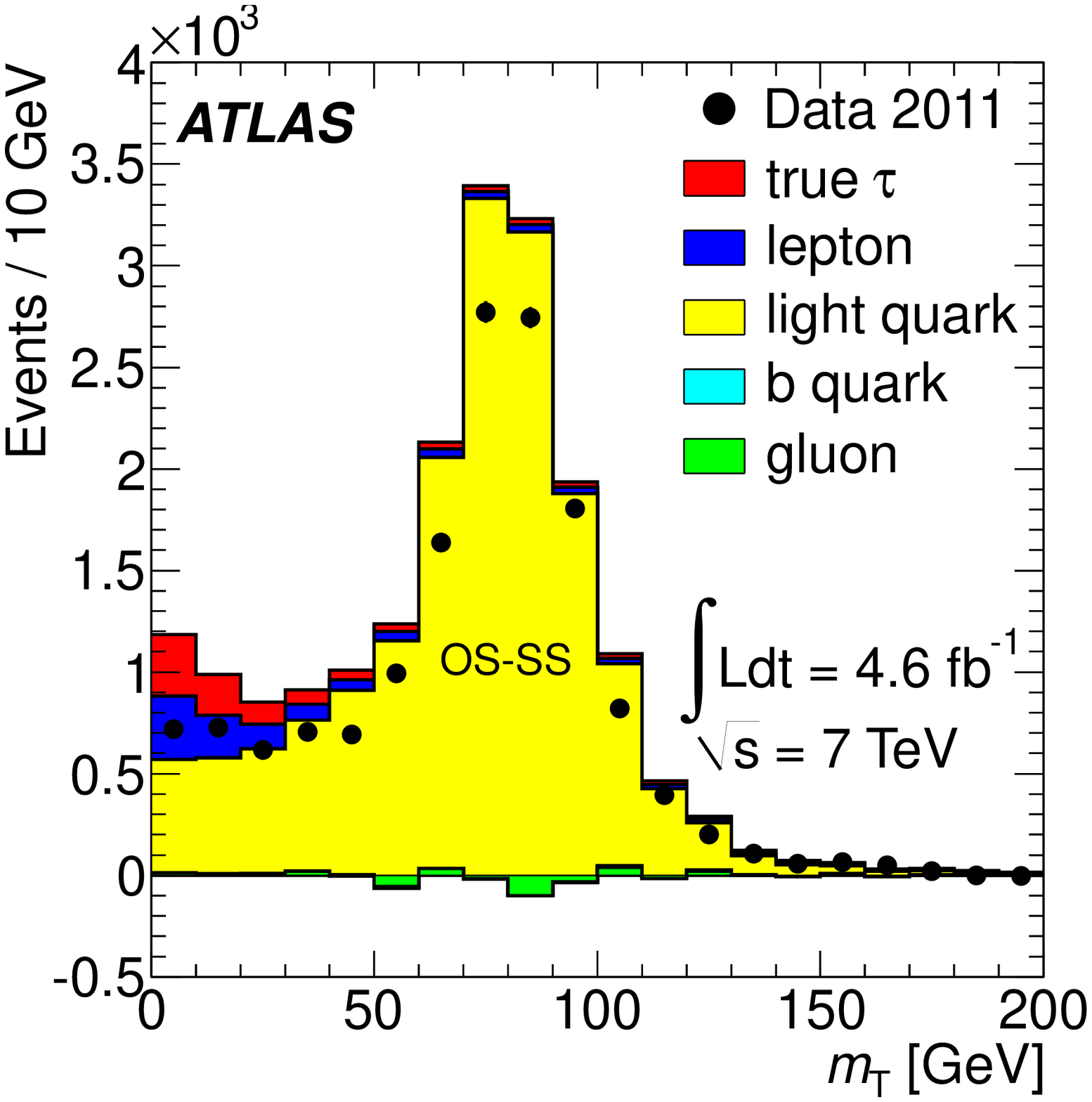}
}
\end{center}
\caption{Distributions of the transverse mass $m_{\mathrm{T}}$ 
for events fulfilling the $W$ + $>$2 jets selection, without the 
requirement $\mt > 30~\mbox{GeV}$. Each colour corresponds to a 
type of generator-level particle matched to a $\tau$ candidate (i.e.\ 
the highest energy particle within a cone of radius $\Delta R = 0.2$ 
around the $\tau$ candidate is considered). The SS events are (a) given 
a weight of $-1$ and (b) subtracted from the OS events. All simulated 
SM processes are considered.
\label{mtOSvsSS}}
\end{figure}

The rate at which light-quark jets are misidentified as $\tau$ 
candidates is derived using a region enriched with $W$ + $>$2 jets 
events\footnote{The leading process in this control region is $gq \to Wq'$.} 
in the data, selected by requiring:
\begin{itemize} 
\item exactly one electron or muon with $E_{\text{T}}$ or $\pt$ larger 
than 25~GeV;
\item at least one $\tau$ candidate; 
\item at least two jets in addition to the $\tau$ candidate(s), none 
of them being $b$-tagged;
\item $\met > 40~\mbox{GeV}$.
\end{itemize}
In order to reduce the contribution from events with a true $\tau$ lepton, 
mostly from $Z$+jets events, a requirement on the transverse mass $\mt$ is 
made:  
\begin{equation}
\mt=
\sqrt{ 2 \pt^l  \ET^{\mathrm{miss}} (1-\cos \Delta\phi_{l,\text{miss}}) }
> 30~\mbox{GeV},
\end{equation}
where $\Delta\phi_{l,\text{miss}}$ is the azimuthal angle between the 
lepton and the direction of the missing momentum. The $W$ + $>$2 jets 
events are classified as OS and SS events using the charges of the 
lepton and the $\tau$ candidate. Figure~\ref{mtOSvsSS} shows the $\mt$ 
distribution for OS, SS and OS-SS events fulfilling the $W$ + $>$2 jets 
selection. This demonstrates the cancellation of heavy-flavour-quark and 
gluon contributions. 

The number of tracks associated to jets misidentified as $\tau$ candidates 
is found to be poorly modelled in simulation. Events in the data tend to 
have fewer $\tau$ candidates with one or three tracks (this explains the 
differences between data and simulation in figure~\ref{mtOSvsSS}). 
In order to correct the $\tau$ candidate selection efficiencies in 
simulation, $\tau$ track multiplicity scale factors are derived using 
OS-SS events fulfilling the $W$ + $>$2 jets selection, and are then 
applied to all jets misidentified as $\tau$ candidates in the simulation: 
$0.71 \pm  0.03$ for 1-track $\tau$ candidates;  
$0.92 \pm  0.03$ for 3-track $\tau$ candidates, 
where the errors are only statistical. 

\begin{figure}[h!]
\begin{center}
\subfigure[]{
\includegraphics[width=0.38\textwidth]{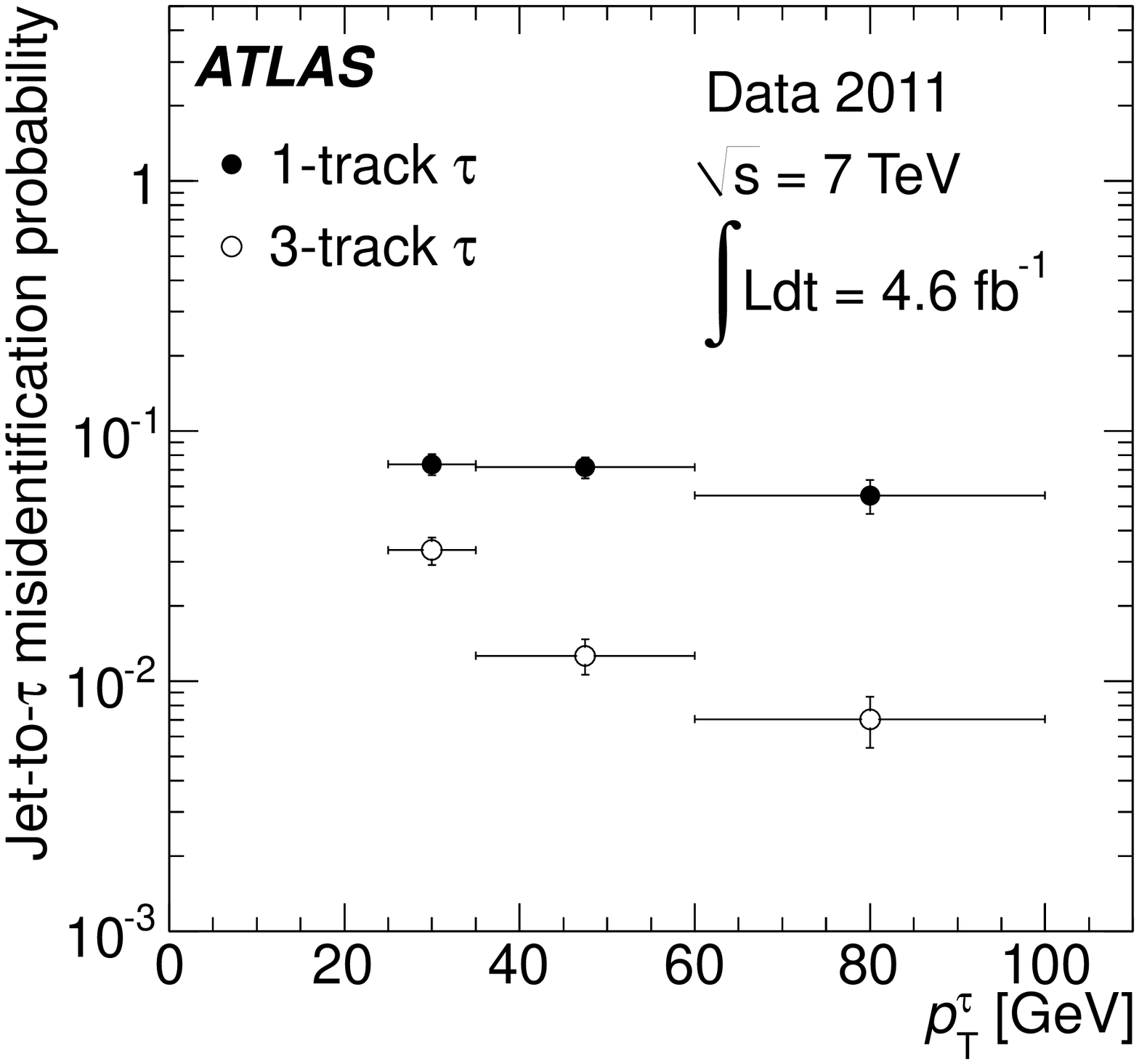}
}
\subfigure[]{
\includegraphics[width=0.38\textwidth]{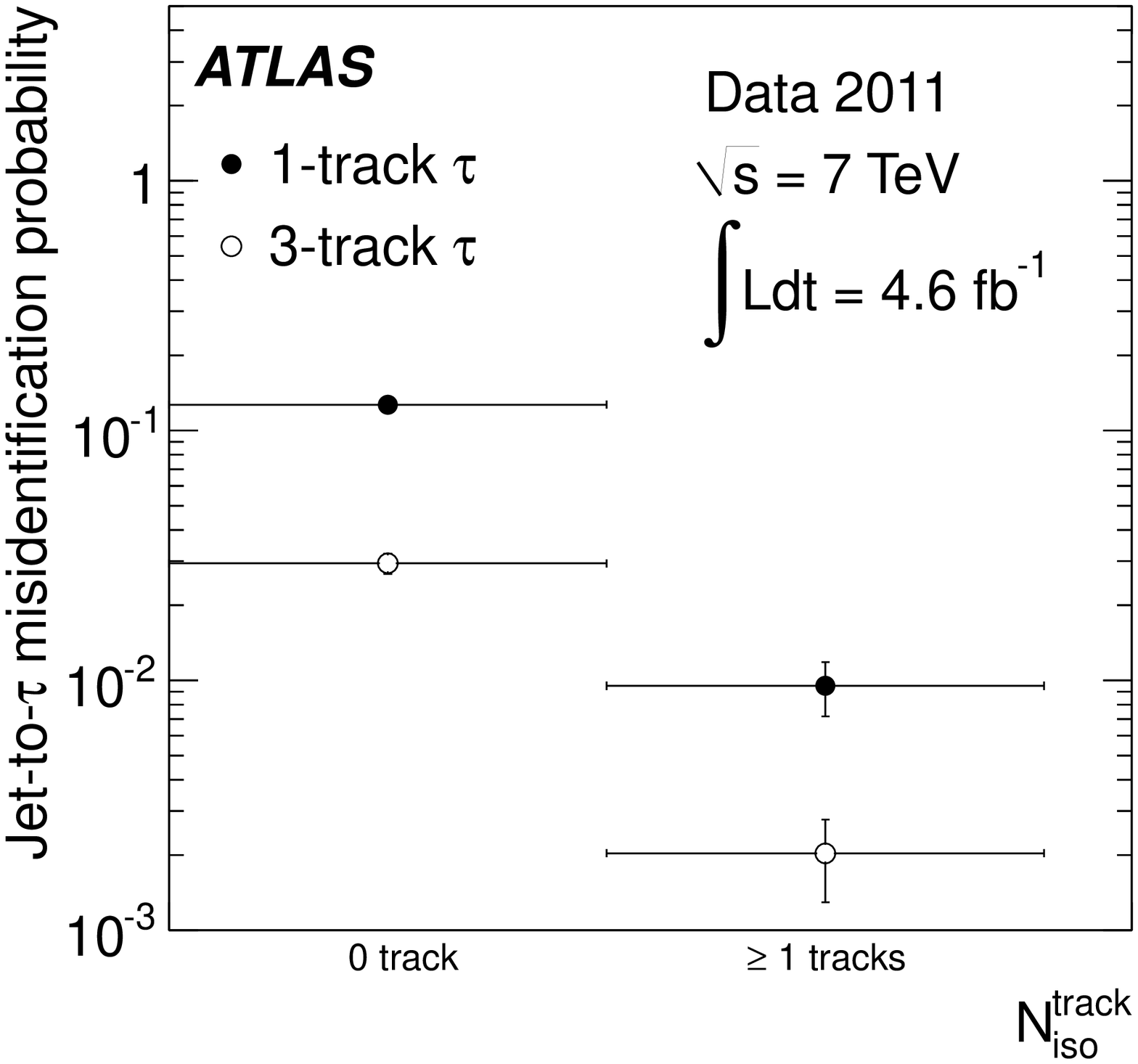}
}
\end{center}
\caption{Probability for a light-quark jet to be misidentified as a 
1-track or 3-track $\tau$ jet, measured in a region enriched with OS-SS 
$W$ + $>$2 jets events in the data, as a function of (a) $\pt^{\tau}$ and 
(b) the number of tracks $N_{\text{track}}^{\text{iso}}$ found within 
$0.2 < \Delta R < 0.4$ of the $\tau$ jet. 
\label{fig:fakerates}}
\end{figure}

The probability for a light-quark jet to be misidentified as a $\tau$ 
jet is measured in the data and is binned in $\pt^{\tau}$, the number of 
associated tracks $N_{\text{track}}^{\tau}$ (one or three), and 
the number of tracks $N_{\text{track}}^{\text{iso}}$ found within 
$0.2 < \Delta R < 0.4$ of the $\tau$ candidate. For each bin, the 
jet$\,\to\tau_{\text{had}}$ misidentification probability is defined 
as the number of objects passing the $\tau$ identification based on 
the likelihood criterion divided by the number prior to requiring 
identification. OS events are given a weight +1 and SS events are 
given a weight $-1$, in both the numerator and denominator of the 
jet$\,\to\tau_{\text{had}}$ misidentification probability. After 
OS-SS subtraction, the selected events mostly contain $\tau$ 
candidates coming from light-quark jets and, to a much lesser extent, 
electrons, muons, and true hadronically decaying $\tau$ leptons. 
Figure~\ref{fig:fakerates} shows the measured values of the 
jet$\,\to\tau_{\text{had}}$ misidentification probability in 
$W$ + $>$2 jets events selected from the data, after OS-SS 
subtraction. These are used to scale all simulated events in 
the signal region. Events fulfilling the requirements listed in 
the beginning of this section, in which the selected $\tau$ object 
originates from a jet (of any type), are weighted by the misidentification 
probabilities. An additional weighting factor (+1 for OS events and $-1$ 
for SS events) is then used to perform the OS-SS subtraction. 

\section{Results}

\subsection{Computation of event yield ratios}

For each of the four final states considered here ($e+\tau_{\text{had}}$, 
$e+\mu$, $\mu+\tau_{\text{had}}$ and $\mu+e$), the OS-SS event yield 
${\cal N}$ can be split into two contributions: from $t\bar{t}$ 
events (where the top quarks may decay into both $bW$ and $bH^+$) 
and from all other SM processes except $t\bar{t} \rightarrow b\bar{b}W^+W^-$. 
The contributions from $t\bar{t}$ events are expressed as a function of the 
cross section $\sigma_{t\bar{t}}$, the integrated luminosity ${\cal L}$, 
the branching fraction $B \equiv {\cal B}(t \rightarrow bH^{+})$, 
as well as the selection efficiencies $\epsilon_{W^+W^-}$, 
$\epsilon_{H^+W^-}$, $\epsilon_{H^-W^+}$ and $\epsilon_{H^+H^-}$ 
for, respectively, $t\bar{t} \to b\bar{b}W^+W^-$, 
$t\bar{t} \to b\bar{b}H^+W^-$, $t\bar{t} \to b\bar{b}H^-W^+$ 
and $t\bar{t} \to b\bar{b}H^+H^-$ events, in each of the four 
final states considered here:  
\begin{eqnarray}
{\cal N} & = & 
\sigma_{t\bar{t}}\times{\cal L}\times
\left[
(1-B)^2\epsilon_{W^+W^-}
+ B(1-B)
\left( 
\epsilon_{H^+W^-} + 
\epsilon_{H^-W^+}
\right)
+ B^2\epsilon_{H^+H^-}
\right] \nonumber \\
& + & {\cal N}_{\text{Others}}~. 
\end{eqnarray} 

In turn, event yield ratios 
are defined as:
\begin{equation}
R_{e} = {{\cal N}(e+\tau_{\text{had}})  \over {\cal N}(e+\mu)} 
\quad 
\text{and} 
\quad 
R_{\mu} = {{\cal N}(\mu+\tau_{\text{had}})  \over {\cal N}(\mu+e)}. 
\end{equation}

The event yields in the $\tau_{\text{had}}$+lepton and dilepton final 
states are summarised in table~\ref{tab:results_taulep} for the 
background-only hypothesis, as well as in the presence of a 130~GeV 
charged Higgs boson in the top quark decay. 
The predicted values in the SM-only hypothesis and the measured 
values of the ratios $R_{e}$ and $R_{\mu}$ are summarised in 
table~\ref{tab:results_ratios}. Note that the event yields for 
dilepton final states become smaller in the presence of a charged 
Higgs boson in top quark decays, despite the fact that a $\tau$ 
lepton decays into an electron or muon more often than a $W$ boson. 
This results from the fact that electrons and muons produced in the 
decay chain $t \to bH^+ \to b\tau\nu \to bl + N\nu$ are, on average, 
softer than those coming from $t \to bW \to bl + N\nu$. 

\begin{table}[h!]
\begin{center}
\begin{tabular}{l|ll}
Sample & \multicolumn{2}{c}{OS-SS event yields} \\
       & \multicolumn{1}{c}{$e+\tau_{\text{had}}$}
       & \multicolumn{1}{c}{$e+\mu$} \\
\hline \hline
Misidentified electrons or muons  & 
$\hspace*{1mm}-0.8 \pm \phantom{0}3.0$ 
                       & 
$\phantom{00}94\phantom{.0} \pm \phantom{0}37$ \\
$W/Z$+jets \& diboson  & 
$\hspace*{1mm}\phantom{+}2.1 \pm \phantom{0}0.9$
                       & 
$\phantom{000}0.7 \pm \phantom{00}0.4$ \\
Single top quark       & 
$\hspace*{1mm}\phantom{+}3.3 \pm \phantom{0}0.8$
                       & 
$\phantom{00}24\phantom{.0} \pm \phantom{00}4$ \\
$t\bar{t}$             & 
$111\phantom{.0} \pm 25$
                       & 
$\phantom{0}980\phantom{.0} \pm 200$ \\
\hline
$\sum$ SM              & 
$116\phantom{.0} \pm 25$
                       & 
$1100\phantom{.0} \pm 210$ \\
\hline
Data         & 
\multicolumn{1}{c}{144} & 
\multicolumn{1}{c}{1247} \\
\hline
$t\bar{t}$ with $t \rightarrow bH^+$ (130~GeV) & 
$\phantom{0}30\phantom{.0} \pm \phantom{0}4$
                               & 
$\phantom{00}27\phantom{.0} \pm \phantom{00}4$ \\
Prediction with signal    & 
$139\phantom{.0} \pm 28$  
                     & 
$1070\phantom{.0} \pm 200\phantom{.0}$ \\

       & \multicolumn{1}{c}{$\mu+\tau_{\text{had}}$}
       & \multicolumn{1}{c}{$\mu+e$}   \\
\hline \hline
Misidentified electrons or muons  & 
$\phantom{00}0.2 \pm \phantom{0}1.0$ 
                       & 
$\phantom{0}74\phantom{.0} \pm \phantom{0}37\phantom{.0}$ \\
$W/Z$+jets \& diboson  & 
$\phantom{00}2.6 \pm \phantom{0}1.6$ 
                       & 
$\phantom{00}0.7 \pm \phantom{00}0.4$ \\
Single top quark       & 
$\phantom{00}4.6 \pm \phantom{0}0.9$ 
                       & 
$\phantom{0}18\phantom{.0} \pm \phantom{00}3\phantom{.0}$ \\
$t\bar{t}$             & 
$131\phantom{.0} \pm 28\phantom{.0}$ 
                       & 
$740\phantom{.0} \pm 150\phantom{.0}$\\
\hline
$\sum$ SM              & 
$138\phantom{.0} \pm 29$ 
                       & 
$830\phantom{.0} \pm 160\phantom{.0}$ \\
\hline
Data         & 
\multicolumn{1}{c}{153} & 
\multicolumn{1}{c}{929} \\
\hline
$t\bar{t}$ with $t \rightarrow bH^+$ (130~GeV) & 
$\phantom{0}35\phantom{.0} \pm \phantom{0}4$
                               & 
$\phantom{0}20\phantom{.0} \pm \phantom{00}3$ \\
Prediction with signal    & 
$166\phantom{.0} \pm 32$
                     & 
$810\phantom{.0} \pm 150$ \\
\end{tabular}
\caption{Expected OS-SS event yields after all selection cuts 
in $\tau_{\text{had}}$+lepton and dilepton channels, compared 
with 4.6 fb$^{-1}$ of ATLAS data. The numbers shown for a 
hypothetical 130~GeV $H^+$ signal correspond to 
${\cal B}(t \rightarrow bH^{+}) = 3\%$. The 
contribution of $t\bar{t} \rightarrow b\bar{b}WW$ events 
to the background is scaled accordingly. 
Statistical and systematic uncertainties are combined.}
\label{tab:results_taulep}
\end{center}
\end{table}

\begin{table}[h!]
\begin{center}
\begin{tabular}{l|cc}
Ratio & $R_e$ & $R_{\mu}$ \\
\hline \hline
SM value & 
$0.105 \pm 0.012$ &
$0.166 \pm 0.017$ \\
Measured value & 
$0.115 \pm 0.010~(\text{stat})$ & 
$0.165 \pm 0.015~(\text{stat})$\\
\end{tabular}
\caption{Predicted (in the SM-only hypothesis) and measured values of the 
event yield ratios $R_{e}$ and $R_{\mu}$. For the values of the ratios 
predicted using simulation, the statistical and systematic uncertainties 
are combined. }
\label{tab:results_ratios}
\end{center}
\end{table}

\begin{figure}[h!]
\begin{center}
\subfigure[]{
\includegraphics[width=0.38\textwidth]{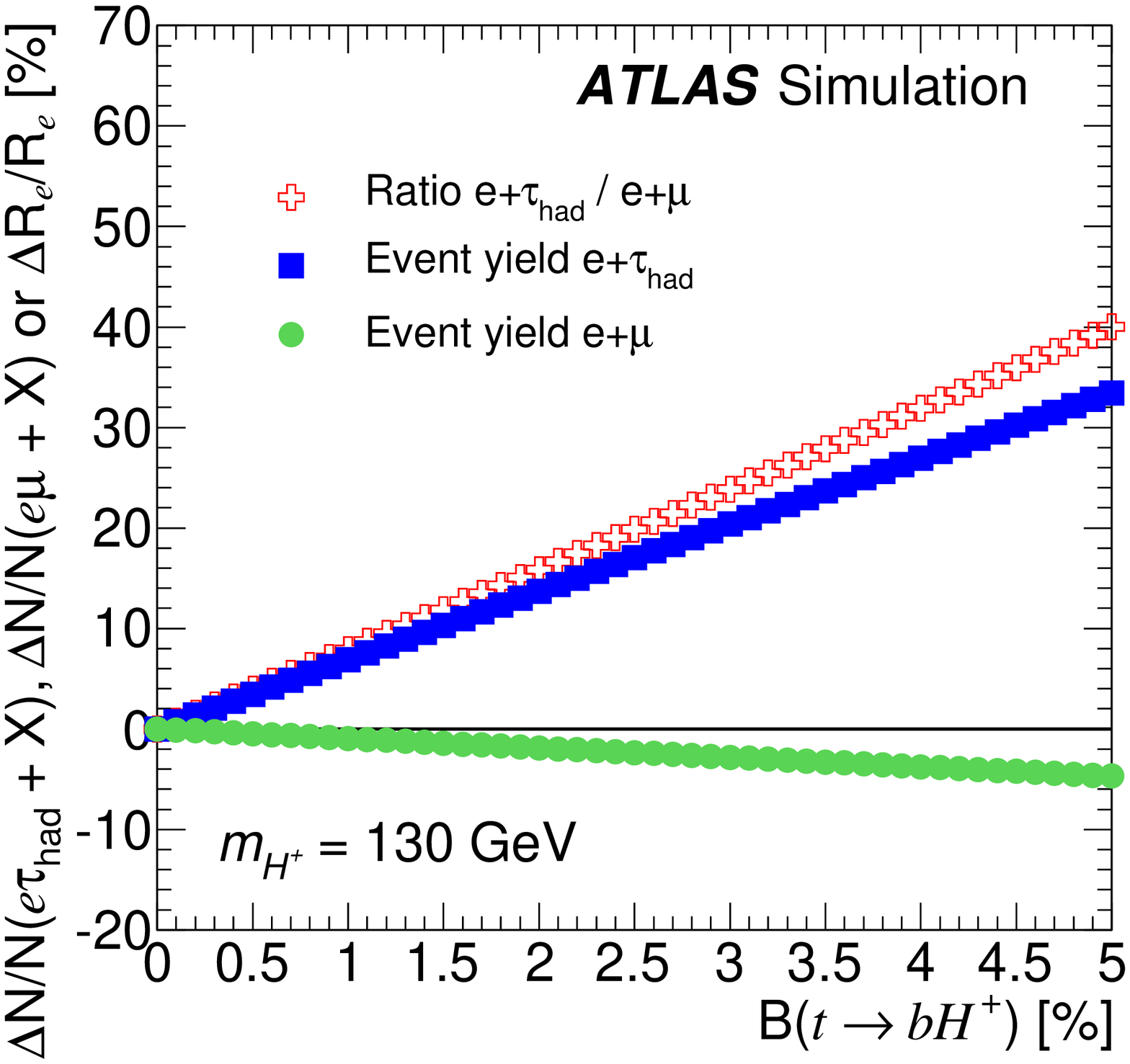}
}
\subfigure[]{
\includegraphics[width=0.38\textwidth]{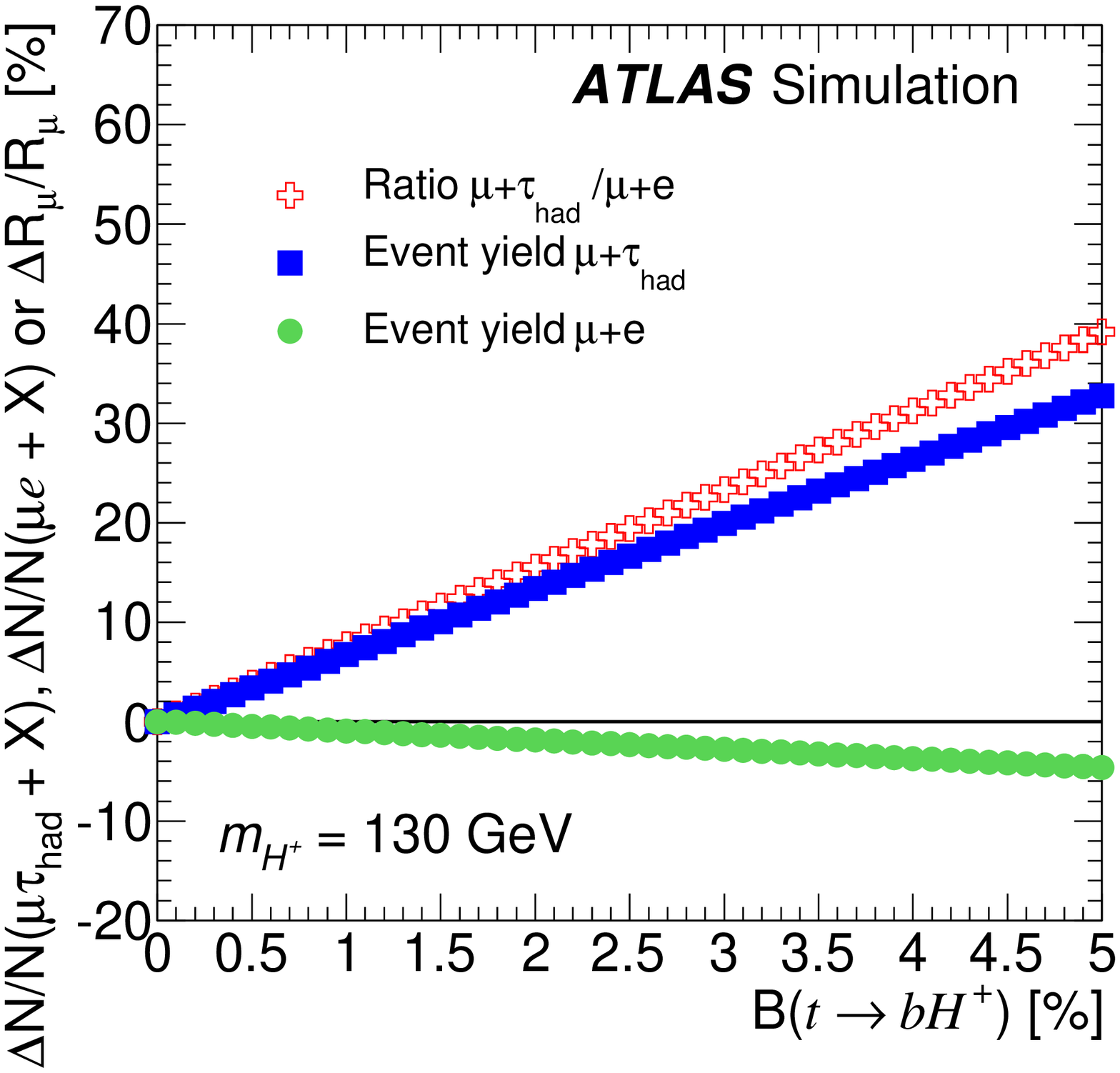}
}
\end{center}
\caption{Relative variation with ${\cal B}(t\to bH^{+})$ of (a) the event 
yields ${\cal N}(e+\tau_{\text{had}})$, ${\cal N}(e+\mu)$ and their ratio, 
as well as (b) ${\cal N}(\mu+\tau_{\text{had}})$, ${\cal N}(\mu+e)$ and 
their ratio, assuming the presence of a 130~GeV charged Higgs boson in 
$t\bar{t}$ events.
\label{ratios130}}
\end{figure}

Figure~\ref{ratios130} shows the variation of the event 
yields ${\cal N}(e+\tau_{\text{had}})$, ${\cal N}(e+\mu)$, 
${\cal N}(\mu+\tau_{\text{had}})$ and ${\cal N}(\mu+e)$ 
with ${\cal B}(t \rightarrow bH^{+})$, 
for a charged Higgs boson mass of 130~GeV. The presence of 
$H^+ \rightarrow \tau\nu$ in a fraction of the top quark 
decays leads to an increase of the number of $t\bar{t}$ 
events with a lepton and a $\tau$ jet. In combination with 
a small decrease of the number of dilepton $t\bar{t}$ events, 
this leads to an increase of the ratios $R_e$ and $R_{\mu}$. 
The sensitivity of this analysis to charged Higgs bosons is 
determined by the rate at which the ratios $R_{e}$ and $R_{\mu}$ 
change with ${\cal B}(t\to bH^{+})$, which depends on the 
selection efficiencies $\epsilon_{H^+W^-}$, $\epsilon_{H^-W^+}$, 
$\epsilon_{H^+H^-}$ and, in turn, on the charged Higgs boson mass.
For $m_{H^+} = 150\,(160)~\mbox{GeV}$, the rate at which the ratios 
$R_{e}$ and $R_{\mu}$ change with ${\cal B}(t\to bH^{+})$ is found 
to be two (five) times smaller than for $m_{H^+} = 130~\mbox{GeV}$.
Indeed, the selection efficiencies $\epsilon_{H^+W^-}$, $\epsilon_{H^-W^+}$, 
$\epsilon_{H^+H^-}$ are reduced for $m_{H^+}$ values in the vicinity 
of $m_{\text{top}}$, because the $b$-jet arising from $t \to bH^+$ 
becomes softer when the mass difference $m_{\text{top}}-m_{H^+}$ is 
smaller. 

\subsection{Systematic uncertainties}
\label{sysunc}

Systematic uncertainties arise from the simulation of the electron and 
muon triggers, from the reconstruction and identification efficiencies 
of the physics objects, as well as from the energy/momentum scale and 
resolution for these objects. In order to assess their impact, the 
selection cuts of this analysis are re-applied after shifting a 
particular parameter by its $\pm 1$ standard deviation uncertainty, 
while other parameters are fixed. 
The largest instrumental systematic uncertainties are for jets. In 
comparison, the systematic uncertainties arising from the reconstruction 
and identification of electrons, muons and $\tau$ jets are small. All 
instrumental systematic uncertainties are propagated to the 
reconstructed $\met$.

The $t\bar{t}$ cross section used in this analysis 
is $\sigma_{t\bar{t}}= 167^{+17}_{-18}$~pb~\cite{hathor}.
To estimate systematic uncertainties arising from the
$t\bar t$ generation and the parton shower model, the acceptance
is computed for $t\bar t$ events produced with either MC@NLO 
interfaced to HERWIG and JIMMY~\cite{jm} for the
hadronisation and the underlying event, or POWHEG~\cite{Frixione:2007vw} 
interfaced to PYTHIA. Systematic uncertainties on initial- and 
final-state radiation are computed using $t\bar{t}$ samples generated with 
AcerMC interfaced to PYTHIA, where the relevant parameters in PYTHIA 
are varied in a range given by current experimental data~\cite{isrfsr}. 
These systematic uncertainties are dominated by the difference in 
modelling the numbers of tracks $N_{\text{track}}^{\tau}$ and 
$N_{\text{track}}^{\text{iso}}$ in, respectively, the core and 
isolation regions of the jets misidentified as $\tau$ candidates. 
The various simulated $t\bar{t}$ samples are reweighted so that 
the $N_{\text{track}}^{\tau}$ and the $N_{\text{track}}^{\text{iso}}$ 
distributions match\footnote{Both variables are reweigthed in a 
correlated way.} before the systematic uncertainties on the $t\bar{t}$ 
generation, the parton shower model, as well as initial- and final-state 
radiation, are evaluated. 

For the signal samples, which are generated with PYTHIA (i.e.\ without 
higher-order corrections), no alternative generator is available, hence 
the systematic uncertainty is set to the relative difference in acceptance 
between $t\bar t$ events generated with MC@NLO interfaced to HERWIG/JIMMY 
and with AcerMC, which is also a leading-order generator, interfaced to 
PYTHIA. For the systematic uncertainty coming from initial- and final-state 
radiation, the same simulated samples as for the SM $t\bar t$ events are 
used. In the evaluation of the systematic uncertainties for the signal 
samples, only $\tau$ jets matched to true hadronically decaying $\tau$ 
leptons in the generated events are considered.

For the backgrounds with misidentified leptons, the largest systematic 
uncertainties arise from the sample dependence: the misidentification 
probabilities are calculated in a control region dominated by 
gluon-initiated events, but later used in a data sample with a 
higher fraction of quark-initiated events. The total systematic 
uncertainty on the backgrounds with misidentified leptons is 
38\% for electron-triggered events and 49\% for muon-triggered 
events. It corresponds to the relative variation of the number of 
events with exactly one trigger-matched lepton and two jets, after 
having considered all systematic uncertainties. The requirement of 
having two $b$-jets in the event does not have a significant impact 
on these systematic uncertainties and neither does the presence of 
a second lepton.

For the estimation of backgrounds with jets misidentified as hadronically 
decaying $\tau$ leptons, the systematic uncertainty on the scale 
factors associated with the number of tracks is determined by varying 
the requirement on the jet multiplicity and the magnitude of the 
subtraction of $\tau$ candidates matched to a true electron, muon 
or $\tau$ lepton in the generated events. This uncertainty is 
7\% for 1-track $\tau$ jets and 11\% for 3-track $\tau$ jets. In addition, 
systematic uncertainties on the jet$\,\to\tau_{\text{had}}$ misidentification 
probability arise from statistical uncertainties due to the limited control 
sample size, the differences between misidentification probabilities computed 
in the region enriched with $W$ + $>$2 jets events and the signal region, 
as well as the small contamination from true $\tau$ leptons (including 
those possibly coming from $H^+ \rightarrow \tau\nu$) in the 
region enriched with $W$ + $>$2 jets events.

Some of the systematic uncertainties above affect the 
$\tau_{\text{had}}$+lepton and dilepton event yields in the 
same manner and, as a result, have a limited impact on 
$R_{e}$ and $R_{\mu}$. 
Systematic uncertainties arising from jets and $\met$ are common 
to all reconstructed events in the simulation, hence they should cancel 
in the ratios $R_{e}$ and $R_{\mu}$. However, due to the use 
of data-driven background estimates and because of the removal of 
geometric overlaps between reconstructed objects, some of these 
systematic uncertainties still have a minor impact. 
In the EL (MU) category, the systematic uncertainties related to the 
trigger-matched electron (muon) are the same for the $e+\tau_{\text{had}}$ 
and $e+\mu$ ($\mu+\tau_{\text{had}}$ and $\mu+e$) events, thereby not 
affecting the predicted value of the ratio $R_{e}$ ($R_{\mu}$). 
Those coming from the reconstructed muon (electron) only affect  
event yields in the denominator, and hence the ratio. Similarly, 
the systematic uncertainties coming from the $\tau$ jets and their 
misidentification probabilities only affect the numerator of $R_{e}$ 
and $R_{\mu}$, hence they do have an impact on the analysis. This is 
also the case for systematic uncertainties on the backgrounds with 
misidentified leptons, which have a larger contribution in the 
dilepton events, i.e.\ on the denominator of $R_{e}$ and $R_{\mu}$.
Table~\ref{tab:sysratios} shows how these ratios (in the SM-only 
hypothesis) change when shifting a particular parameter by its $\pm 1$ 
standard deviation uncertainty. 
 
\begin{table}[h!]
\begin{center}
\begin{tabular}{l|c|c}
\centering
Systematic uncertainty & $\Delta R_{e}$ & $\Delta R_{\mu}$ \\
\hline \hline
Integrated luminosity & 0.3\%  & 0.3\% \\
\hline \hline
Electron trigger efficiency & 0.1\%  & N/A \\
Electron reco. and ID efficiencies & 0.2\%  & 1.9\% \\
Electron energy resolution & 0.1\%  & $<$0.1\% \\
Electron energy scale & 0.1\%  & 0.3\% \\
\hline
Muon trigger efficiency & N/A  & 0.1\% \\
Muon reco. and ID efficiencies & 1.0\%  & 0.1\% \\
Muon momentum resolution & $<$0.1\%  & $<$0.1\% \\
Muon momentum scale & 0.1\%  & $<$0.1\% \\
\hline
$\tau$ ID efficiency & 3.9\%  & 3.9\% \\
$\tau$ energy scale & 2.9\%  & 3.0\% \\
$\tau$ mis-ID (data-driven): number of associated tracks & 
2.1\%  & 2.1\% \\
$\tau$ mis-ID (data-driven): true $\tau_{\text{had}}$ contamination & 
0.2\%  & 0.2\% \\
$\tau$ mis-ID (data-driven): $H^+$ signal contamination & 
0.6\%  & 0.6\% \\
$\tau$ mis-ID (data-driven): event environment & 
1.3\%  & 1.2\% \\
$\tau$ mis-ID (data-driven): statistical uncertainties & 
3.3\%  & 3.2\% \\
$\tau$ mis-ID (data-driven): electron veto uncertainties & 
0.6\%  & 0.3\% \\
\hline
$b$-tagging & 1.9\%  & 2.3\% \\
Jet vertex fraction & 0.1\%  & 0.4\% \\
Jet energy resolution & 0.4\%  & $<$0.1\% \\
Jet energy scale & 0.7\%  & 0.5\% \\
Jet reconstruction efficiency & 0.1\%  & 0.4\% \\
\hline
$\met$  & 0.3\%  & 0.1\% \\
\hline \hline
$t\bar{t}$: cross section & 0.7\% & 0.6\% \\ 
$t\bar{t}$: generator and parton shower & 5.7\% & 4.4\% \\ 
$t\bar{t}$: initial- and final-state radiation & 3.6\% & 3.7\% \\ 
\hline
Backgrounds with misidentified leptons & 3.5\% & 4.3\% \\
\hline \hline
Total (added in quadrature) & 10.3\% & 10.1\% \\
\end{tabular}
\caption{Relative variation of the ratios $R_{e}$ and $R_{\mu}$ in the 
SM-only hypothesis after shifting a particular parameter by its 
$\pm 1$ standard deviation uncertainty.}
\label{tab:sysratios}
\end{center}
\end{table}
   
\subsection{Exclusion limits}
\label{sec:limit}

\begin{figure}[h!]
\begin{center}
\subfigure[Using $R_e$]{
\includegraphics[width=0.475\textwidth]{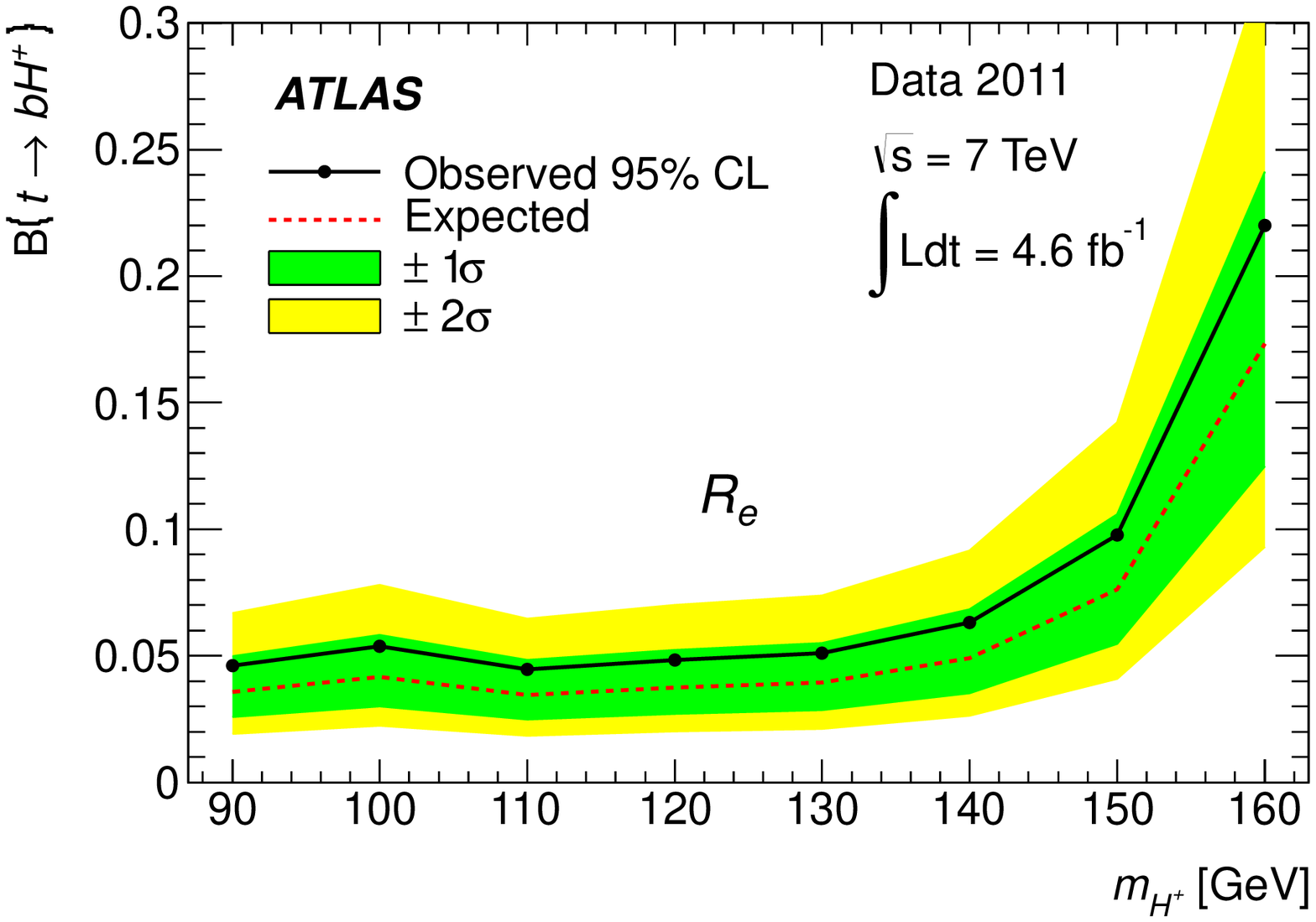}
}
\subfigure[Using $R_{\mu}$]{
\includegraphics[width=0.475\textwidth]{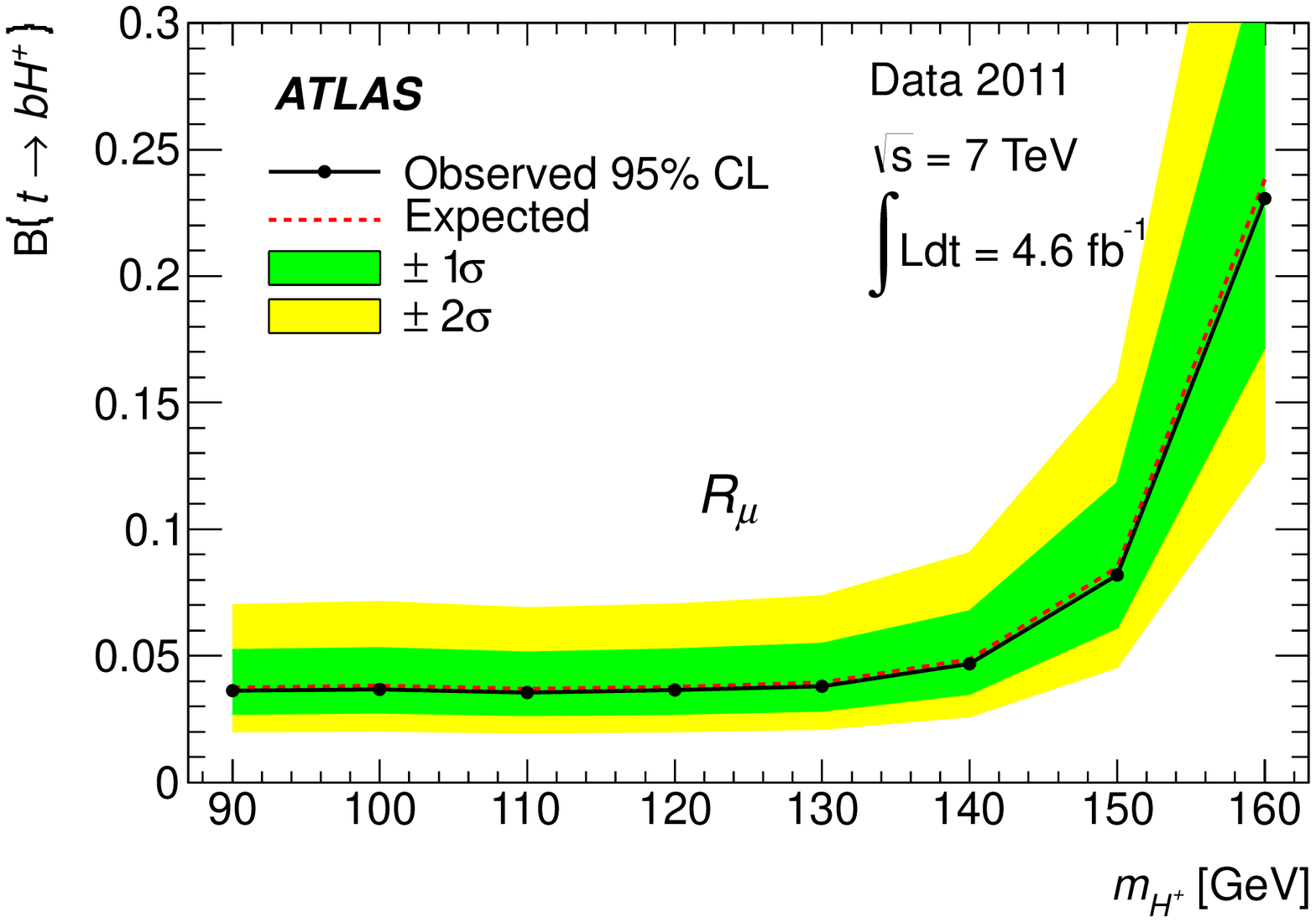}
}
\end{center}
\caption{Upper limits on ${\cal B}(t\rightarrow bH^+)$ derived from 
the event yield ratios (a) $R_e$ and (b) $R_\mu$, as a function of 
the charged Higgs boson mass, obtained for an integrated luminosity 
of 4.6~fb$^{-1}$ and with the assumption 
${\cal B}( H^+ \rightarrow \tau\nu) = 1$. The solid line in the figure is 
used to denote the observed 95\% CL upper limits, while the dashed line 
represents the expected exclusion limits. The green and 
yellow regions show the 1$\sigma$ and 2$\sigma$ error bands.
\label{limits-ratios}}
\end{figure}

\begin{figure}[h!]
\begin{center}
\includegraphics[width=0.63\textwidth]{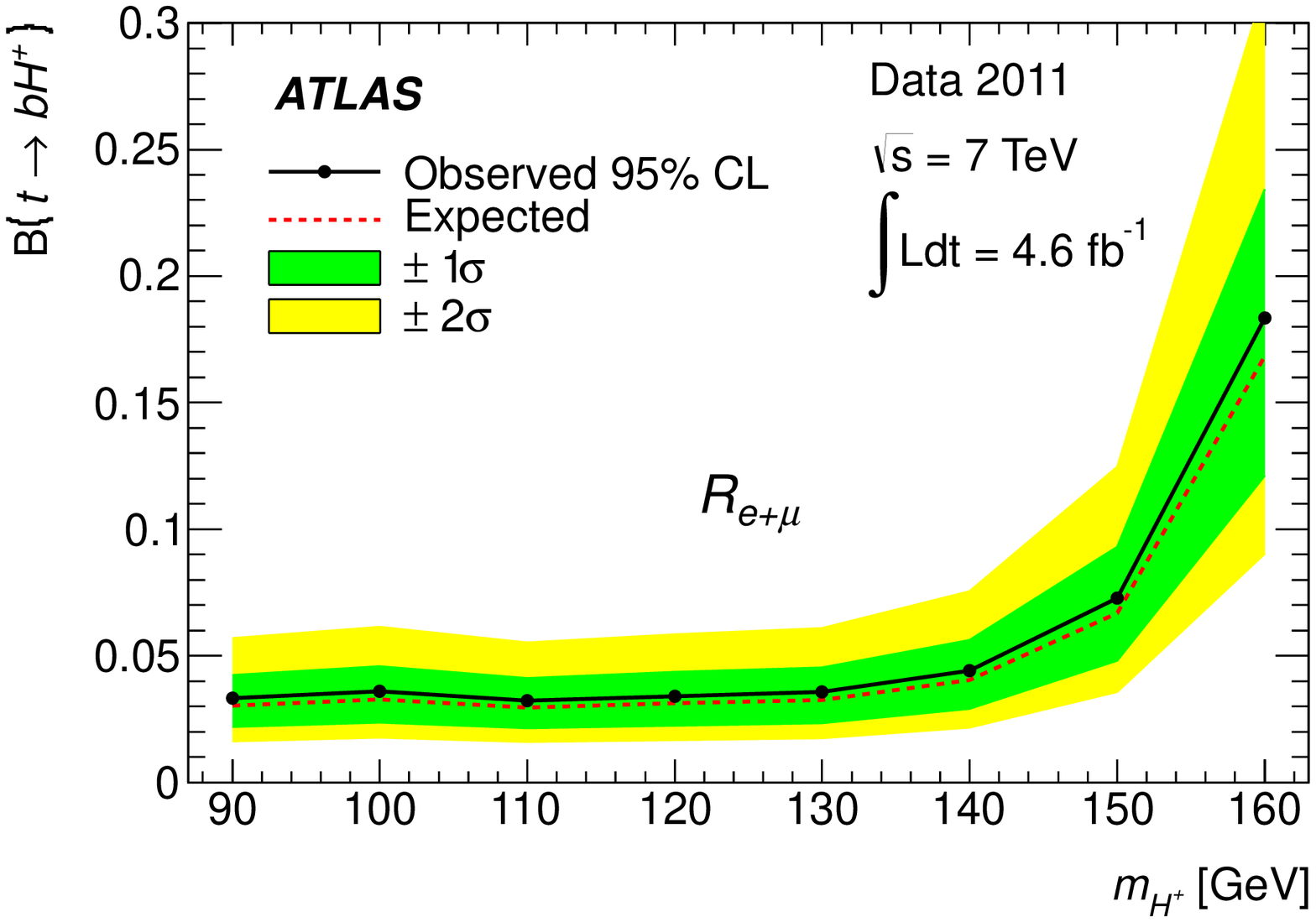}
\end{center}
\caption{Upper limits on ${\cal B}(t\rightarrow bH^+)$ derived from 
the event yield ratio $R_{e+\mu}$, as a function of the charged Higgs 
boson mass, obtained for an integrated luminosity of 4.6~fb$^{-1}$ and 
with the assumption ${\cal B}( H^+ \rightarrow \tau\nu) = 1$. 
The solid line in the figure is used to denote the observed 95\% CL 
upper limits, while the dashed line represents the expected exclusion 
limits. The green and yellow regions show the 
1$\sigma$ and 2$\sigma$ error bands.
\label{limits-comb}}
\end{figure}

To test the compatibility of the data with the background-only 
or the signal+background hypotheses, a profile likelihood 
ratio~\cite{Cowan:2010js} is used with $R_{e}$ and $R_{\mu}$ 
as the discriminating variables. The systematic uncertainties 
are incorporated via nuisance parameters, and the one-sided profile 
likelihood ratio, $\tilde{q}_{\mu}$, is used as a test statistic. 
No significant deviation from the SM prediction is observed in 
\LUMI of data. Exclusion limits are set on the branching fraction 
${\cal B}(t \to bH^+)$ by rejecting the signal hypothesis at the 95\% 
confidence level (CL) using the $CL_s$ procedure~\cite{Read:2002hq}. 
These limits are based on the asymptotic distribution of the test 
statistic~\cite{Cowan:2010js}. They are first set for electron-triggered 
and muon-triggered events separately (see figure~\ref{limits-ratios}), and 
then using a global event yield ratio $R_{e+\mu}$ defined as:
\begin{equation}
R_{e+\mu} = {{{\cal N}(e+\tau_{\text{had}})
            +   {\cal N}(\mu+\tau_{\text{had}})} \over 
              {{\cal N}(e+\mu)+{\cal N}_{\text{OR}}(\mu+e)}},
\end{equation}
where ${\cal N}_{\text{OR}}(\mu+e)$ is the event yield in the $\mu+e$ 
channel after removing the dilepton events that simultaneously 
fire a single-electron trigger and a single-muon trigger, as those 
already appear in ${\cal N}(e+\mu)$. The fraction of dilepton events 
common to the $\mu+e$ and $e+\mu$ final states is about 42\% in the 
data. Using this global event yield ratio, upper limits in the range 
3.2\%--4.4\% can be placed on ${\cal B}(t\rightarrow bH^+)$ for 
charged Higgs boson masses in the range 90--140~GeV, as shown in 
figure~\ref{limits-comb} and table~\ref{tab:limits-comb}.

\begin{figure}[h!]
\begin{center}
\includegraphics[width=0.63\textwidth]{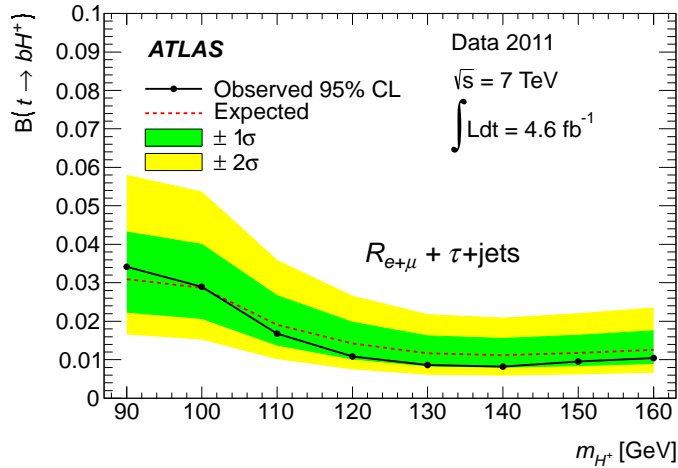}
\end{center}
\caption{Upper limits on ${\cal B}(t\rightarrow bH^+)$ derived 
from the transverse mass distribution of $\tau_{\text{had}}$+jets 
events in ref.~\cite{ourpaper} and the event yield ratio $R_{e+\mu}$, as 
a function of the charged Higgs boson mass, obtained for an 
integrated luminosity of 4.6~fb$^{-1}$ and with the assumption 
${\cal B}( H^+ \rightarrow \tau\nu) = 1$. 
The solid line in the figure is 
used to denote the observed 95\% CL upper limits, while the dashed line 
represents the expected exclusion limits. The green and 
yellow regions show the 1$\sigma$ and 2$\sigma$ error bands.
\label{taujets-comb}}
\end{figure}

In a previously published search for charged Higgs bosons~\cite{ourpaper}, 
based on the data collected in 2011 with ATLAS, upper limits on  
${\cal B}(t\rightarrow bH^+)$ were derived using various distributions 
of discriminating variables in $\tau_{\text{had}}$+jets, 
$\tau_{\text{had}}$+lepton and lepton+jets final states. 
The most sensitive channel was $\tau_{\text{had}}$+jets, 
except for low values of $m_{H^+}$. 
A new set of combined upper limits on ${\cal B}(t\rightarrow bH^+)$ 
is derived, using the transverse mass distribution of 
$\tau_{\text{had}}$+jets events from ref.~\cite{ourpaper} 
and the global event yield ratio $R_{e+\mu}$, as shown in 
figure~\ref{taujets-comb} and table~\ref{tab:taujets-comb}.
Since a lepton veto is applied for charged Higgs boson searches in 
$\tau_{\text{had}}$+jets final states, there is no correlation between 
such events and those selected in this study to determine the 
event yield ratios.
With this combination of upper limits, charged Higgs bosons 
can be excluded for values of the branching fraction 
${\cal B}(t\rightarrow bH^+)$ 
larger than 0.8\% to 3.4\%, for $m_{H^+}$ between 90~GeV 
and 160~GeV. These exclusion limits represent an improvement 
with respect to those published in ref.~\cite{ourpaper}.

\begin{table}[h!]
\begin{center}
\small
\begin{tabular}{l|cccccccc} 
$m_{H^+}$ (GeV) & 90 & 100 & 110 & 120 & 130 & 140 & 150 & 160 \\
\hline \hline
95\% CL observed & & & & & & & & \\ 
(expected) limit on & 
3.3\% & 3.6\% & 3.2\% & 3.4\% & 3.6\% & 4.4\% & 7.3\% & 18.3\% \\
${\cal B}(t\rightarrow bH^+)$ using& 
(3.1\%) & (3.3\%) & (3.0\%) & (3.1\%) & (3.3\%) & (4.0\%) & (6.7\%) & (16.8\%)  \\
the ratio $R_{e+\mu}$& & & & & & & & \\ 
\end{tabular}
\normalsize
\caption{Observed (expected) 95\% CL upper limits on
${\cal B}(t\rightarrow bH^+)$ derived from the event yield 
ratio $R_{e+\mu}$, as a function of the charged Higgs boson mass, 
obtained for an integrated luminosity of 4.6~fb$^{-1}$ and with the 
assumption that ${\cal B}(H^+ \rightarrow \tau\nu) = 1$. }
\label{tab:limits-comb}
\end{center}
\end{table}  

\begin{table}[h!]
\begin{center}
\small
\begin{tabular}{l|cccccccc} 
$m_{H^+}$ (GeV) & 90 & 100 & 110 & 120 & 130 & 140 & 150 & 160 \\
\hline \hline
95\% CL observed & & & & & & & & \\ 
(expected) limit on & 
3.4\% & 2.9\% & 1.7\% & 1.1\% & 0.9\% & 0.8\% & 1.0\% & 1.1\% \\
${\cal B}(t\rightarrow bH^+)$ using& 
(3.1\%) & (2.8\%) & (1.9\%) & (1.4\%) & (1.2\%) & (1.1\%) & (1.2\%) & (1.2\%)  \\
$R_{e+\mu}$ and $\tau_{\text{had}}$+jets & & & & & & & & \\ 
\hline 
95\% CL observed & & & & & & & & \\ 
(expected) limit& 
4.8\% & 3.4\% & 2.1\% & 1.3\% & 1.1\% & 1.0\% & 1.1\% & 1.0\% \\
in ref.~\cite{ourpaper}& 
(4.2\%) & (3.5\%) & (2.5\%) & (1.9\%) & (1.5\%) & (1.3\%) & (1.2\%) & (1.3\%)  \\
\end{tabular}
\normalsize
\caption{Observed (expected) 95\% CL upper limits on
${\cal B}(t\rightarrow bH^+)$ derived using $\tau_{\text{had}}$+jets 
events in ref.~\cite{ourpaper} and the ratio $R_{e+\mu}$, 
as a function of the charged Higgs boson mass, obtained for 
an integrated luminosity of 4.6~fb$^{-1}$ and assuming that 
${\cal B}(H^+ \rightarrow \tau\nu) = 1$. The exclusion limits 
published in ref.~\cite{ourpaper} are also shown for comparison 
purposes.}
\label{tab:taujets-comb}
\end{center}
\end{table}

\begin{figure}[h!]
\begin{center}
\includegraphics[width=0.5\textwidth]{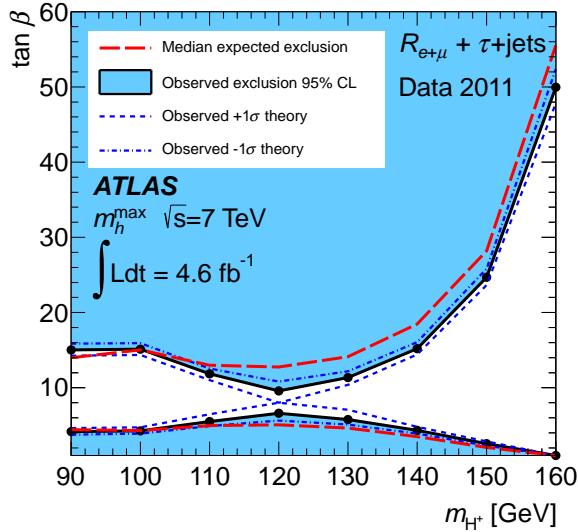}
\end{center}
\caption{Limits for charged Higgs boson production from top quark
decays in the $m_{H^+}$-$\tan\beta$ plane, derived using 
$\tau_{\text{had}}$+jets 
events in ref.~\cite{ourpaper} and the ratio $R_{e+\mu}$, in the 
context of the $m_{h}^{\text{max}}$ scenario of the MSSM. The $1\sigma$ 
band around the observed limit (dashed lines) shows the theoretical 
uncertainties. Values below $\tan\beta=1$, where the calculations in 
the MSSM become non-perturbative, are not considered, as the results 
become unphysical.
\label{mssm2}}
\end{figure}

In figure~\ref{mssm2}, the combined limit on 
${\cal B}(t\rightarrow bH^+) \times {\cal B}(H^+\rightarrow\tau\nu)$ 
is interpreted in the context of the $m_h^{\mathrm{max}}$ 
scenario~\cite{Carena:2002qg} of the MSSM. 
The following relative theoretical uncertainties on ${\cal B}(t \to bH^+)$ 
are considered~\cite{Heinemeyer:1998yj,Dittmaier:2012vm}: 
5\% for one-loop electroweak corrections missing from the calculations, 
2\% for missing two-loop QCD corrections, and about 1\% (depending on 
$\tan \beta$) for $\Delta_b$-induced uncertainties, where $\Delta_b$ is a 
correction factor for the running $b$-quark mass~\cite{Carena:1999py}. 
These uncertainties are added linearly, as recommended by 
the LHC Higgs cross-section working group~\cite{Dittmaier:2012vm}.

Assuming that the boson recently discovered at the 
LHC~\cite{higgsATLAS,higgsCMS} is one of the neutral 
MSSM Higgs bosons, only a certain region in the 
$m_{H^+}$-$\tan\beta$ plane is still allowed for 
a given scenario~\cite{arXiv:1112.3026}. If the new 
boson is the lightest neutral MSSM Higgs boson ($h^0$), 
it would imply $\tan \beta > 3$ and $m_{H^+}>155~\mbox{GeV}$. 
However, the allowed region depends strongly on MSSM parameters which, 
on the other hand, do not affect the charged Higgs boson production 
and decay significantly. Thus, by adjusting these MSSM parameters, the 
region in which the Higgs boson mass can take a value of about 125~GeV 
can be changed significantly, while the ATLAS exclusion region shown here 
is relatively stable with respect to these changes. Should the recently 
discovered boson instead be the heavier CP-even Higgs boson ($H^0$), the 
additional constraint from $m_{H^0} \simeq 125~\mbox{GeV}$ only leads to 
an upper limit of roughly $m_{H^+}<150$~GeV, with suppressed couplings 
for $h^0$. 
If the recently discovered particle is an MSSM Higgs boson, 
excluding a low-mass charged Higgs boson would thus imply that it 
is the lightest neutral state $h^0$.


\section{Conclusions}
\label{sec:conclusion}

Charged Higgs bosons have been searched for in \ttbar\ events, in the decay 
mode $t \to bH^+$ followed by $H^+ \to \tau\nu$. A total of \LUMI of $pp$ 
collision data at $\sqrt{s}=7~\mbox{TeV}$, recorded in 2011 with the 
ATLAS experiment at the LHC, is used. 
Event yield ratios are measured in the data and compared to the predictions 
from simulations, between electron-triggered $e+\tau_{\text{had}}$ and 
$e+\mu$ events, and between muon-triggered $\mu+\tau_{\text{had}}$ and 
$\mu+e$ events, in order to search for a violation of lepton universality 
in $t\bar{t}$ events. This method reduces the impact of several systematic 
uncertainties in the analysis. Data-driven methods and 
simulation are employed to estimate the number of background events. 
The observed data are found to be in agreement with the SM predictions. 
Assuming ${\cal B}(H^+ \to \tau\nu)=100\%$, upper limits at the 95\% 
confidence level in the range 3.2\%--4.4\% have been placed on the 
branching fraction ${\cal B}(t \to bH^{+})$ for charged Higgs boson 
masses in the range 90--140~GeV. 
For charged Higgs boson masses below 110~GeV, this analysis improves 
the previously published limits on ${\cal B}(t\rightarrow bH^+)$, based 
on direct searches for charged Higgs bosons in $t\bar{t}$ decays using 
the lepton+jets, $\tau_{\text{had}}$+jets and $\tau_{\text{had}}$+lepton 
final states.
When the results of the present analysis are combined with 
the results from the search for charged Higgs bosons in $t\bar{t}$ 
decays using the $\tau_{\text{had}}$+jets final state~\cite{ourpaper}, 
upper limits on ${\cal B}(t\rightarrow bH^+)$ are set in the range 
0.8\%--3.4\%, for $m_{H^+}$ between 90~GeV and 160~GeV. 


\section{Acknowledgements}

We thank CERN for the very successful operation of the LHC, as well as the
support staff from our institutions without whom ATLAS could not be
operated efficiently.

We acknowledge the support of ANPCyT, Argentina; YerPhI, Armenia; ARC,
Australia; BMWF and FWF, Austria; ANAS, Azerbaijan; SSTC, Belarus; CNPq 
and FAPESP, Brazil; NSERC, NRC and CFI, Canada; CERN; CONICYT, Chile; 
CAS, MOST and NSFC, China; COLCIENCIAS, Colombia; MSMT CR, MPO CR and 
VSC CR, Czech Republic; DNRF, DNSRC and Lundbeck Foundation, Denmark; 
EPLANET, ERC and NSRF, European Union; IN2P3-CNRS, CEA-DSM/IRFU, France; 
GNSF, Georgia; BMBF, DFG, HGF, MPG and AvH Foundation, Germany; GSRT and 
NSRF, Greece; ISF, MINERVA, GIF, DIP and Benoziyo Center, Israel; INFN, 
Italy; MEXT and JSPS, Japan; CNRST, Morocco; FOM and NWO, Netherlands; 
BRF and RCN, Norway; MNiSW, Poland; GRICES and FCT, Portugal; 
MERYS (MECTS), Romania; MES of Russia and ROSATOM, Russian Federation; 
JINR; MSTD, Serbia; MSSR, Slovakia; ARRS and MVZT, Slovenia; DST/NRF, 
South Africa; MICINN, Spain; SRC and Wallenberg Foundation, Sweden; SER, 
SNSF and Cantons of Bern and Geneva, Switzerland; NSC, Taiwan; TAEK, Turkey; 
STFC, the Royal Society and Leverhulme Trust, United Kingdom; DOE and NSF, 
United States of America.

The crucial computing support from all WLCG partners is acknowledged
gratefully, in particular from CERN and the ATLAS Tier-1 facilities at
TRIUMF (Canada), NDGF (Denmark, Norway, Sweden), CC-IN2P3 (France),
KIT/GridKA (Germany), INFN-CNAF (Italy), NL-T1 (Netherlands), PIC (Spain),
ASGC (Taiwan), RAL (UK) and BNL (USA) and in the Tier-2 facilities
worldwide.


\bibliographystyle{JHEP}
\bibliography{HIGG-2012-22}

\clearpage 
\begin{flushleft}
{\Large The ATLAS Collaboration}

\bigskip

G.~Aad$^{\rm 48}$,
T.~Abajyan$^{\rm 21}$,
B.~Abbott$^{\rm 111}$,
J.~Abdallah$^{\rm 12}$,
S.~Abdel~Khalek$^{\rm 115}$,
A.A.~Abdelalim$^{\rm 49}$,
O.~Abdinov$^{\rm 11}$,
R.~Aben$^{\rm 105}$,
B.~Abi$^{\rm 112}$,
M.~Abolins$^{\rm 88}$
% ...
\bigskip
\\
% ...
$^{11}$ Institute of Physics, Azerbaijan Academy of Sciences, Baku, Azerbaijan\\
$^{12}$ Institut de F{\'\i}sica d'Altes Energies and Departament de F{\'\i}sica de la Universitat Aut{\`o}noma de Barcelona and ICREA, Barcelona, Spain\\
% ...
$^{21}$ Physikalisches Institut, University of Bonn, Bonn, Germany\\
% ...
$^{48}$ Fakult{\"a}t f{\"u}r Mathematik und Physik, Albert-Ludwigs-Universit{\"a}t, Freiburg, Germany\\
$^{49}$ Section de Physique, Universit{\'e} de Gen{\`e}ve, Geneva, Switzerland\\
% ...
$^{88}$ Department of Physics and Astronomy, Michigan State University, East Lansing MI, United States of America\\
% ...
$^{105}$ Nikhef National Institute for Subatomic Physics and University of Amsterdam, Amsterdam, Netherlands\\
% ...
$^{111}$ Homer L. Dodge Department of Physics and Astronomy, University of Oklahoma, Norman OK, United States of America\\
$^{112}$ Department of Physics, Oklahoma State University, Stillwater OK, United States of America\\
% ...
$^{115}$ LAL, Universit{\'e} Paris-Sud and CNRS/IN2P3, Orsay, France\\
% ...
\end{flushleft}


\end{document}